\documentstyle[buckow]{article}
\def\be{\begin{equation}}
\def\ee{\end{equation}}
\def\bea{\begin{eqnarray}}
\def\eea{\end{eqnarray}}
\def\barray{\begin{array}}
\def\earray{\end{array}} 
\def\nn{\nonumber \\}

\def\part{\partial}

\def\tfrac#1#2{{\textstyle{#1\over #2}}}

\def\STr{\mbox{STr}}

\def\incl{\mbox{i}}

\def\tb{\tilde b}
\def\cD{{\cal D}}
\def\cP{{\cal P}}

\def\a9{{a9}}
\def\mn{{\mu\nu}}
\def\mnr{{\mu\nu\rho}}

\def\al{\alpha}
\def\la{\lambda} 
\def\td{\tilde}

\begin {document}
\rightline{\small DCPT-02/03, FFUOV-02/01}  

\def\titleline{
Chern-Simons Couplings for Dielectric F-Strings in Matrix String Theory
}
\def\authors{
Dominic Brecher\1ad, Bert Janssen\1ad and Yolanda Lozano\2ad }
\def\addresses{
\1ad
Centre for Particle Theory, Department of Mathematical Sciences, \\
University of Durham, South Road, Durham DH1 3LE, 
United Kingdom. \\ 
{\tt email: dominic.brecher, bert.janssen@durham.ac.uk} \\~ \\
\2ad Departamento de F{\'\i}sica, Universidad de Oviedo \\ 
Avda. Calvo Sotelo 18, 33007 Oviedo, Spain.\\
{\tt e-mail:  yolanda@string1.ciencias.uniovi.es} \\
}

\def\abstracttext{
We compute the non-abelian couplings in the Chern-Simons action for a set of coinciding
fundamental strings in both the type IIA and type IIB Matrix string
theories. Starting from Matrix theory in a weakly curved background,
we construct the linear couplings of closed string
fields to type IIA Matrix strings.  Further dualities give a
type IIB Matrix string theory and a type IIA theory of Matrix strings
with winding.\footnote{Talk given by Bert Janssen.} }
\large
\makefront
\section{Introduction}

It is well-known that a collection of D-branes under the 
influence of a background R-R field strength can undergo an 
``expansion'' into a single higher-dimensional D-brane. This is 
the so-called dielectric effect, a first analysis of which was performed in
\cite{emparan:97}, at the level of the
abelian theory relevant to the description of the single 
expanded D-brane.  
It was some years later that the complementary\footnote{See \cite{myers}
for a discussion on the ranges of validities of the two descriptions.} 
description from the point of view
of the lower-dimensional multiple branes was provided~\cite{myers}.  
From this perspective, the expansion takes place because the 
embedding coordinates of the multiple branes are matrix-valued,
and give rise to new non-abelian couplings in the
combined Born-Infeld-Chern-Simons action \cite{TVR, myers}.

There is much evidence that the dielectric effect should also exist 
for fundamental 
strings, both from the abelian analysis \cite{emparan:97} and from the supergravity
perspective \cite{emparan2}. However, one would further like to provide the 
complementary description
from the point of view of the fundamental strings. Since, 
from the strings' perspective, the dielectric effect should be due to matrix-valued 
coordinates, one is led to a consideration of Matrix string theory \cite{DVV}.
Moreover, matrix string 
theory is equivalent to type IIA superstring theory
in the light-cone gauge, together with extra degrees of freedom
representing D-brane states. Therefore it is the appropriate framework
in which to study systems of multiple fundamental strings expanding
into higher-dimensional D-branes (see also \cite{pedro}).  Starting from 
Matrix theory in a weakly
curved background, we construct the linear couplings of closed string
fields to type IIA Matrix strings.   Further dualities give a
type IIB Matrix string theory and a type IIA theory of Matrix strings
with winding. From these actions we identify the couplings that are
responsible for the dielectric effect. 
 
This letter is a summary of the results published in \cite{BJL}.

\section{Matrix Theory and D0-branes}
 
Our starting point is the Chern-Simons action for D0-branes with linear couplings 
to the R-R background fields, as given in \cite{TVR}:
\bea
S_{{\rm linear}} = \tfrac{1}{R} \int dt ~\STr \left\{ 
C^{(1)}_{\mu} I_0^{\mu} + C^{(3)}_{\mnr}
I_2^{\mnr} + \tfrac{1}{60} C^{(5)}_{\mu_1 \ldots \mu_5} I_4^{\mu_1 \ldots \mu_5}
+ \tfrac{1}{336} C^{(7)}_{\mu_1 \ldots \mu_7} I_6^{\mu_1 \ldots \mu_7} \right\},
\label{eqn:D0}
\eea
where $\STr$ denotes the symmetrised trace and $R$ the radius of the 
eleventh dimension. These couplings can be derived from the 
Matrix theory couplings to linear background fields,
as shown in \cite{KT}. The indices $\mu, \nu = 0,\ldots,9$,  
the currents $I_p$ couple to the R-R $(p+1)$-form potentials $C^{(p+1)}$ and
the potentials $C^{(5)}$ and $C^{(7)}$ have been rescaled relative 
to~\cite{TVR}\footnote{Note that there should also be couplings to $C^{(9)}$, 
but these are not determined by the analysis of~\cite{TVR}.}. The currents
appearing in (\ref{eqn:D0}) are given in terms of the dimensional
reduction of the Born-Infeld field strength,
\be
F_{0i} = -F^{0i} = \partial_t X^i \equiv \dot{X}^i, \qquad F_{ij} = F^{ij} =
\frac{i}{\beta} [X^i, X^j],
\label{eqn:BI}
\ee
where $i,j=1,\ldots,9$ and  $\beta = 2\pi l_P^3/R$.
We will not give the explicit expression but refer to \cite{TVR}. It 
can be shown \cite{myers, BJL} that (\ref{eqn:D0}) agrees with the linear order
expansion of the multiple D0-brane theory 
\cite{TVR, myers}
\bea
S = T_0 \int dt ~ \STr \left\{ P \left[ e^{ i (\incl_X
\incl_X)/\la} \left( \sum C^{(n)} \right) \right] \right\},
\label{eqn:chern-simons}
\eea
with $ \lambda = 2\pi \al'$. Here the interior multiplication is defined as
$( \incl_X \Sigma )_{\mu_1 \ldots \mu_{p}} = X^i \Sigma_{i \mu_1 \ldots
\mu_{p}}$. This type of contraction of the embedding scalars with the
R-R potentials are the origin of the dielectric 
effect, as explained in \cite{myers}.

\section{Multiple D-strings}
To construct the Matrix string theory of \cite{DVV}, one compactifies Matrix theory
on a
circle in, say, the $x^9$ direction and performs a T-duality transformation.
Taking now $i,j = 1, \ldots,8$, and
denoting the dual coordinate by $\hat{x}$, the world volume fields
transform as~\cite{taylor}
\bea
F^{9i} = \tfrac{i}{\beta} [X^9,X^i] \longrightarrow 
\tfrac{1}{2\pi\hat{R}_9} \int d\hat{x} \frac{\la}{\beta} D_{\hat{x}} X^i, 
\hspace{1cm}
F^{09} = -\dot{X}^9 \longrightarrow -\tfrac{1}{2\pi \hat{R}_9} \int
d\hat{x} \la
\dot{A}_{\hat{x}},
\label{eqn:t-duality2}
\eea
where $\hat{R}_9 = \al'/R_9$ is the radius of the dual circle. 
Turning to the linear action (\ref{eqn:D0}), we must consider
T-duality applied to both the world volume and background fields.  
As far as the currents are
concerned, we just have to consider the transformation of
the Born-Infeld field strength (\ref{eqn:BI}) under
T-duality, as given by
(\ref{eqn:t-duality2}) above.  This
is a simple re-writing, for the results we refer to \cite{TVR}. 

To linear order, the action of T-duality on the R-R fields is:
\be
C^{(p)}_{a_1 \ldots a_{p-1} 9} \ \longleftrightarrow \ C^{(p-1)}_{a_1 \ldots
a_{p-1}},
\label{eqn:t-duality_bgd2}
\ee
where $a,b = 0, \ldots, 8$. A simple application of these rules gives
\bea
S_{{\rm linear}} &=& \tfrac{1}{2\pi R \hat{R}_9} \int dt d\hat{x} ~\STr \left\{
 C^{(0)} I_0^9 + C^{(2)}_{a9} I_0^a + 3 C^{(2)}_{ab} I_2^{ab9} + C^{(4)}_{abc9}
I_2^{abc} + \tfrac{1}{12} C^{(4)}_{a_1 \ldots a_4} I_4^{a_1 \ldots a_4 9} 
\right.\nn
&&\left. \qquad \qquad 
+ \tfrac{1}{60} C^{(6)}_{a_1 \ldots a_5 9} I_4^{a_1
\ldots a_5}
+ \tfrac{1}{48} C^{(6)}_{a_1 \ldots a_6} I_6^{a_1 \ldots a_6 9}
+ \tfrac{1}{336} C^{(8)}_{a_1 \ldots a_7 9} I_6^{a_1 \ldots a_7}
\right\}.
\label{eqn:d1_linear}
\eea
By construction, the multiple D0-brane action (\ref{eqn:chern-simons}) 
is covariant under T-duality, so the T-dual of the 
D0-brane action (\ref{eqn:D0}) should be equivalent to the linearised
version of the D-string action.  With the $I$s as in \cite{TVR}, it is
easy to see that the R-R terms can indeed be written as \cite{myers}
\be
S = T_1 \int \STr \left\{ P \left[ e^{ i (\incl_X
\incl_X)/\la} \left( \sum C^{(n)} \right) \right] \wedge e^{\la F} \right\},
\ee
where $F_{09} = \dot{A}_{\hat x}$.\footnote{We should note that only half of
the terms necessary to form the pullback of $C^{(8)}$ are present in
the linear action (\ref{eqn:d1_linear}).  The missing
terms come from the missing $C^{(9)}$ coupling in the D0-brane action 
(\ref{eqn:D0}).}

\section{Matrix string theory in type IIA}
Having constructed the $(1+1)$-dimensional theory for multiple D-strings, we
are now in a position to perform the so-called 9-11 flip, 
which acts on the background fields as
\be
\begin{array}{lll}
C^{(0)} \longrightarrow -C^{(1)}_9, 
& C^{(2)}_{ab} \longrightarrow b_{ab}, 
& C^{(2)}_{a9} \longrightarrow -h_{a9},
\\ \\ 
C^{(4)}_{a_1 \ldots a_4} \longrightarrow C^{(5)}_{a_1 \ldots a_49},
&C^{(4)}_{abc9} \longrightarrow C^{(3)}_{abc}, 
& C^{(6)}_{a_1\ldots a_6} \longrightarrow N^{(7)}_{a_1 \ldots a_6 9},
\\ \\
C^{(6)}_{a_1 \ldots a_5 9} \longrightarrow {\tilde b}_{a_1 \ldots a_5 9},
&C^{(8)}_{a_1 \ldots a_8} \longrightarrow -C^{(9)}_{a_1 \ldots a_8 9},
& C^{(8)}_{a_1 \ldots a_7 9} \longrightarrow -C^{(7)}_{a_1 \ldots
a_7},
\label{eqn:911_h}
\end{array}
\ee
and will give us the type IIA Matrix string
theory action in a linear background.  Here, $b_\mn$, $\tb_{\mu_1..\mu_6}$ 
and $N^{(7)}_{\mu_1 \ldots \mu_7}$ are the NS-NS 2-form, its Hodge dual and 
the field that couples minimally to the type IIA Kaluza-Klein monopole, respectively.
We take the view here that the currents are invariant under the 9-11 flip, 
since in the flat space case 
it does not change the worldvolume fields~\cite{DVV}.

Performing the transformations (\ref{eqn:911_h}) on
the linear action (\ref{eqn:d1_linear}), one finds
\[
S = \tfrac{1}{2\pi} \int d\tau d\sigma \tfrac{\al'}{R^2}
~\STr \left\{
- C^{(1)}_9 I_0^9 - h_{a9} I_0^a + 3 b_{ab} I_2^{ab9} + C^{(3)}_{abc} I_2^{abc}
+ \tfrac{1}{12} C^{(5)}_{a_1 \ldots a_4 9} I_4^{a_1 \ldots a_4 9}  \right.
\]
\be
\left.
+ \tfrac{1}{60} \td{b}_{a_1 \ldots a_5 9} I_4^{a_1 \ldots a_5}
+ \tfrac{1}{48} N^{(7)}_{a_1 \ldots a_69} I_6^{a_1 \ldots a_6 9}
- \tfrac{1}{336} C^{(7)}_{a_1 \ldots a_7} I_6^{a_1 \ldots a_7}
\right\}.
\label{eqn:linear_mst}
\ee
Here, we have defined the dimensionless world sheet coordinates
\be
\sigma = \tfrac{1}{\hat{R}_9}\ \hat{x}^9, \qquad 
\tau = \tfrac{R}{\al'} \ t.
\label{eqn:rescalings}
\ee
Writing the currents $I$ in terms of the dimensionless
quantities $\tau$ and $\sigma$, this is the action describing Matrix
string theory in a weakly curved background \cite{schiappa, BJL}. 

In particular, if we set the Born-Infeld field to zero, we find
for the R-R three-form couplings:
\be
S_{C^{(3)}} = \tfrac{i}{4\pi g_s}  \int d\tau d\sigma
~\STr \left\{ \tfrac{\sqrt{\al'}}{R} \tfrac{1}{\sqrt{2}}C^{(3)}_+ 
+ \tfrac{\sqrt{\al'}}{R} \tfrac{1}{\sqrt{2}}C^{(3)}_- 
+ \dot{X}^i C^{(3)}_i \right\},
\label{C^3}
\ee
where we have defined
\be
C^{(3)}_\mu = [X^j, X^i] C^{(3)}_{\mu ij},
\ee
and made use of light-cone coordinates.  This coupling has been given
before in \cite{schiappa}. We showed however
in \cite{BJL} that extra $C^{(3)}$ terms arise if one takes into account the
couplings coming from the Born-Infeld part of the action. A dielectric
solution involving only the couplings in (11) was given in \cite{schiappa}
which we have interpreted in \cite{BJL} in terms of gravitational waves
expanding into a transverse D2-brane.

It can be seen from (\ref{eqn:linear_mst}) that as regards the R-R
5-form potential, only the terms of the form $C^{(5)}_{a_1...a_4 9}$
contribute. Setting the Born-Infeld vector to 
zero, the remaining 5-form R-R field couplings can be written as:
\be
S_{C^{(5)}} = \tfrac{i}{4\pi g_s} \tfrac{R}{\sqrt{\al'}} \int d\tau d\sigma
~\STr \left\{ \tfrac{\sqrt{\al'}}{R} C^{(5)}_{+-} + \dot{X}^i C^{(5)}_{+i} 
                  -\dot{X}^i C^{(5)}_{-i} \right\},
\ee
where we have defined
\be
C^{(5)}_\mn = [X^k, X^j] DX^i C^{(5)}_{ijk\mn}.
\ee
Similarly, the $C^{(7)}$ couplings only have contributions involving
terms of the form $C^{(7)}_{a_1 ... a_7}$. These can be written as 
\be
S_{C^{(7)}} = \tfrac{i}{96\pi g_s^3} \tfrac{R^2}{\al'} \int d\tau d\sigma
~\STr \left\{ \tfrac{\sqrt{\al'}}{R}\tfrac{1}{\sqrt{2}} C^{(7)}_{+} +
              \tfrac{\sqrt{\al'}}{R}\tfrac{1}{\sqrt{2}} C^{(7)}_{-} +   
                \dot{X}^i C^{(7)}_{i} \right\},
\ee
with
\be
C^{(7)}_\mu = [X^n, X^m][X^l, X^k][X^j, X^i] C^{(7)}_{ijklmn\mu}.
\ee
Let us also consider for the sake of completeness the couplings to
$\tb^{(6)}$ and $N^{(7)}$.  As for $C^{(5)}$, only the terms with a
9-component appear in the action. The
$\tb^{(6)}$ couplings can be written as:
\be
S_{\tb^{(6)}} = \tfrac{1}{16\pi g_s^2} \tfrac{R}{\sqrt{\al'}} \int d\tau d\sigma
~\STr \left\{ \tfrac{\sqrt{\al'}}{R} \tb^{(6)}_{+-} + \dot{X}^i \tb^{(6)}_{+i} 
                                                 -\dot{X}^i \tb^{(6)}_{-i}
\right\},
\ee
where 
\be
\tb^{(6)}_\mn = [X^l, X^k][X^j,X^i]  \tb^{(6)}_{ijkl\mn},
\ee
and the $N^{(7)}$ couplings are given by
\be
S_{N^{(7)}} = -\tfrac{1}{16\pi g_s^2} \tfrac{R^2}{\al'} \int d\tau d\sigma
~\STr \left\{ \tfrac{\sqrt{\al'}}{R} N^{(7)}_{+-} - \dot{X}^i N^{(7)}_{+i} 
                                                 +\dot{X}^i N^{(7)}_{-i} \right\},
\ee
where 
\be
N^{(7)}_\mn = [X^m, X^l][X^k,X^j]DX^i  N^{(7)}_{ijklm\mn}.
\ee
Note that the couplings of the different fields occur at a different order of the
expansion 
para\-meter $R/\sqrt{\al'}$. One should look at these couplings when 
studying fundamental strings expanding into D4-, D6- or NS5-branes, or
Kaluza-Klein monopoles.

\section{Multiple IIB F-strings}

To describe fundamental strings in the type IIB theory, we perform
another T-duality in the $x^9$ direction, as in (\ref{eqn:t-duality_bgd2}).  
As before,
we assume that the world volume fields do not change.  
The linear action (\ref{eqn:linear_mst}) becomes
\[
S_{{\rm linear}}^{{\rm IIB}} = \tfrac{1}{2\pi}\int d\tau d\sigma \tfrac{\al'}{R^2} 
~\STr \left\{ 
 b_{a9} I_0^a - C^{(0)} I_0^9   
+ C^{(4)}_{abc9} I_2^{abc} + 3 b_{ab} I_2^{ab9}
\right.
\]
\be
\left. 
+ \tfrac{1}{12}C^{(4)}_{a_1...a_4} I_4^{a_1...a_4 9}
+ \tfrac{1}{60}\tb_{a_1...a_5 9} I_4^{a_1...a_5} 
+ \tfrac{1}{48}\tb_{a_1\ldots a_6} I_6^{a_1...a_6 9} 
- \tfrac{1}{336} C^{(8)}_{a_1...a_7 9} I_6^{a_1...a_7} \right\},
\label{F1B}
\ee
which should describe Matrix strings in the IIB theory.
Note that precisely the same action is obtained if one applies the
S-duality rules to the D1-brane action
(\ref{eqn:d1_linear}), although  it is not
clear \emph{a priori} how such an S-duality should be 
done directly, since we are dealing with non-abelian fields.  
The action above can be rewritten in a more convenient form, filling in the
expressions for the currents:
\[
S = \tfrac{1}{2\pi} \int d\tau d\sigma ~\STr \left\{ P
\left[ 
\tfrac{\al'}{R^2} b^{(2)} + \tfrac{g_s\sqrt{\al'}}{R} C^{(0)}\wedge
F + \tfrac{i\sqrt{\al'}}{g_s R} (\incl_X \incl_X) C^{(4)} + i (\incl_X
\incl_X)b^{(2)} \wedge F 
\right. \right.
\]
\be
 - \tfrac{1}{2 g_s^2}(\incl_X \incl_X)^2\tb^{(6)}
- \tfrac{R}{2g_s\sqrt{\al'}}(\incl_X \incl_X)^2 C^{(4)} \wedge F  
+ \tfrac{iR}{6 g_s^3\sqrt{\al'}}(\incl_X \incl_X)^3 C^{(8)}
- \tfrac{iR^2}{6 g_s^2 \al'}(\incl_X \incl_X)^3 \tb^{(6)}\wedge F 
              \vphantom{\frac{i\sqrt{\al'}}{g_s R}}          \Bigr] \Bigr\}.
\label{F1B2}
\ee
Again we  note that only half of the terms necessary to form the pullback 
of $C^{(8)}$ are present in the linear action. 
Some of the above couplings have been given before in \cite{yolanda}. In \cite{BJL}
it was shown that the $C^{(4)}$ coupling gives rise to a solution of
F-strings expanding into a D3-brane. 

\section{Multiple F-strings with winding in type IIA}

Type IIA F-strings with winding number can be obtained from IIB strings 
by performing a T-duality transformation in a direction transverse to 
the IIB strings. Calling $z$ the T-duality direction and $a=(0, i)$, 
where now $i= 1, ...,7$,  the linear action that is obtained from 
(\ref{F1B}) is given by:
\[
S_{{\rm linear}}^{{\rm IIA}} = \tfrac{1}{2 \pi} \int d\tau d\sigma
\tfrac{\al'}{R^2} ~\STr \left\{
b_{a9} I_0^a
+ h_{z9} I_0^z - C^{(1)}_z I_0^9  + C^{(5)}_{abc9z} I_2^{abc} - 3
C_{ab9} I_2^{abz} + 3 b_{ab} I_2^{ab9}  \right.
\]
\[
- 6 h_{az} I_2^{az9}
+ \tfrac{1}{60} N^{(7)}_{a_1 ... a_59z} I_4^{a_1 ... a_5}
+ \tfrac{1}{12} \tb_{a_1 ... a_4z9} I_4 ^{a_1 ... a_4z}\nn
+ \tfrac{1}{12} C^{(5)}_{a_1 ... a_4z} I_4^{a_1 ... a_49}
+ \tfrac{1}{3} C^{(3)}_{abc} I_4^{abcz9}
\]
\be
\left.
- \tfrac{1}{336} C^{(9)}_{a_1 ... a_79z} I_6^{a_1 ... a_7}
+ \tfrac{1}{48} C^{(7)}_{a_1 ... a_69} I_6^{a_1 ... a_6z} +
\tfrac{1}{48}N^{(7)}_{a_1 ... a_6z} I_6^{a_1 ... a_69}
+ \tfrac{1}{8} \tb_{a_1 ... a_5z} I_6^{a_1 ... a_5z9} \right\} .
\label{F1A*}
\ee
The direction $z$ in which the T-duality is performed appears as an isometry
direction in the transverse space of the strings. We denote the 
corresponding Killing vector as
$k^\mu = \delta_z^\mu$, 
$k_\mu = \eta_{z\mu} + h_{z\mu}$.
In a manner similar to the Kaluza-Klein 
monopole, the non-abelian strings do not see this special direction, the
embedding scalar $X^z$ is not a degree of freedom of the strings, but is 
transformed under T-duality into a world volume scalar $\omega$ 
\footnote{This world volume scalar forms an invariant field strength with $b_{az}$
(see \cite{yolanda} for the details), and can therefore be associated to
fundamental strings wrapped around the isometry direction $z$, which
themselves end on the Matrix strings.}.

The action (\ref{F1A*}) can be written in a covariant way as a gauged sigma 
model, where gauge covariant derivatives $\cD_\alpha X^\mu$ are used to gauge away
the  embedding scalar corresponding to the isometry direction \cite{BJO}:
\be
\cD_\alpha X^\mu = D_\alpha X^\mu - k_\rho D_\alpha X^\rho k^\mu,
\ee
with $\alpha = \sigma, \tau$. These gauge covariant derivatives reduce to the
standard covariant derivatives $D_\alpha X^\mu$ for $\mu \neq z$ and are zero for 
$\mu = z$. The pull-backs in the action of the F-strings with winding are
constructed from these gauge covariant derivatives. For example,
\bea
\cP \left[ b^{(2)} \right] 
              &=& b_\mn \cD X^\mu \cD X^\nu dt dx
\eea 
Filling in the expressions for the currents,
we can write the Chern-Simons action as:
\[
S_{{\rm linear}} = \tfrac{1}{2\pi} \int d\tau d\sigma
~\STr \left\{ 
\cP \left[ \vphantom{\frac{1}{2}}
-\tfrac{ g_s\sqrt{\al'}}{R} \incl_k C^{(1)} \wedge  F 
+ \tfrac{\al'}{R^2} b^{(2)} 
- \tfrac{\sqrt{\al'}}{R} k^{(1)} \wedge D\omega + 
i (\incl_{[X,\omega]}) k^{(1)} \wedge F \right. \right.
\]
\[
+  \tfrac{i\sqrt{\al'}}{g_sR} (\incl_{[X, \omega]}) C^{(3)}
+ \tfrac{i}{g_s} (\incl_X \incl_X)  C^{(3)}\wedge D\omega  
+ \tfrac{R}{g_s\sqrt{\al'}} (\incl_X \incl_X)(\incl_{[X, \omega]})
C^{(3)} \wedge F - \tfrac{i\sqrt{\al'}}{g_sR} (\incl_X \incl_X)  \incl_k C^{(5)} 
\]
\[
+ \tfrac{R}{2g_s\sqrt{\al'}} (\incl_X \incl_X)^2 \incl_k C^{(5)} \wedge F 
+ \tfrac{1}{g_s^2} (\incl_{[X, \omega]})(\incl_X \incl_X) \incl_k
\tb^{(6)} +  \tfrac{R}{2 g_s^2\sqrt{\al'}}(\incl_X \incl_X)^2 \incl_k
\tb^{(6)}\wedge D\omega
\]
\[
+ \tfrac{R^2}{2 g_s\al'} (\incl_{[X, \omega]})(\incl_X \incl_X)^2
          \incl_k \tb^{(6)}\wedge F - \tfrac{1}{2 g_s^2} (\incl_X \incl_X)^2 \incl_k
N^{(7)} 
- \tfrac{iR^2}{6 g_s^2\al'} (\incl_X \incl_X)^3 \incl_k N^{(7)} \wedge F 
\]
\be
\left. \left.  - 
\tfrac{iR^2}{2g_s^3\al'}  (\incl_{[X, \omega]}) (\incl_X \incl_X)^3 C^{(7)} 
+ \tfrac{R}{6 g_s^3\sqrt{\al'}}  (\incl_X \incl_X)^3\incl_k C^{(9)} \right] \right\},
\label{F1A*CS}
\ee
where we have introduced the following types of interior multiplication:
\bea
(i_k \Sigma)_{\mu_1...\mu_p} = k^\rho \Sigma_{\rho\mu_1...\mu_p}
                              =  \Sigma_{z\mu_1...\mu_p},
\hspace{1.5cm}
 (\incl_{[X, \omega]} \Sigma)_{\mu_1...\mu_p} = [X^i, \omega]
\Sigma_{i\mu_1...\mu_p}.
\eea 
Again only half of the terms to form the pullback of $C^{(7)}$ and  $C^{(9)}$ 
are present in the linear action (\ref{F1A*}). Some of the couplings
in (\ref{F1A*CS}) have been derived earlier in \cite{yolanda}. It can be shown 
\cite{BJL} that the $C^{(5)}$ couplings lead to a solution of F-strings expanding 
into a D4-brane that is smeared in the $z$ direction and the $C^{(3)}$ couplings 
to an unstable solution of F-strings expanding into a cylindrical D2-brane.


\end{document}